\documentclass[sigconf,screen]{acmart}
\setcopyright{acmlicensed}
\acmDOI{10.1145/3650212.3680341}
\acmYear{2024}
\copyrightyear{2024}
\acmISBN{979-8-4007-0612-7/24/09}
\acmConference[ISSTA '24]{Proceedings of the 33rd ACM SIGSOFT International Symposium on Software Testing and Analysis}{September 16--20, 2024}{Vienna, Austria}
\acmBooktitle{Proceedings of the 33rd ACM SIGSOFT International Symposium on Software Testing and Analysis (ISSTA '24), September 16--20, 2024, Vienna, Austria}
\acmSubmissionID{issta24main-p543-p}
\received{2024-04-12}
\received[accepted]{2024-07-03}

\AtBeginDocument{%
  \providecommand\BibTeX{{%
    \normalfont B\kern-0.5em{\scshape i\kern-0.25em b}\kern-0.8em\TeX}}}

\usepackage{algorithm}
\usepackage{tabularx}
\usepackage{algpseudocode}
\usepackage{tcolorbox}
\usepackage{graphicx}
\usepackage{textcomp}
\usepackage{xcolor}
\usepackage{array}
\usepackage{subcaption}
\usepackage{enumitem}
\usepackage{longtable}
\usepackage{colortbl}
\usepackage{xcolor}
\usepackage{tikz}
\usepackage{listings}
\usepackage{graphicx}
\usepackage{threeparttable}
\usepackage{multirow}
\usepackage{tikz}
\usepackage{mdframed}
\usepackage{soul}
\usepackage{hyperxmp}
\usepackage{hyperref}

\newmdenv[backgroundcolor=yellow]{highlighted}

\newcommand{\commentdbw}[1]{{\color{blue} \sf (DW: #1)}}

\newcommand{\ignore}[1]{}
\usepackage{xspace}
\newcommand{\Name}{\textsc{ReBL}\xspace}

\title{Feedback-Driven Automated Whole Bug Report Reproduction for Android Apps}

\author{Dingbang Wang}
\affiliation{%
  \institution{University of Connecticut}
  \country{USA}
  }
\email{dingbang.wang@uconn.edu}

\author{Yu Zhao}
\affiliation{%
  \institution{University of Cincinnati}
  \country{USA}
  }
\email{zhao3y3@ucmail.uc.edu}

\author{Sidong Feng}
\affiliation{%
  \institution{Monash University}
  \country{Australia}
  }
\email{sidong.feng@monash.edu}

\author{Zhaoxu Zhang}
\affiliation{%
  \institution{University of Southern California}
  \country{USA}
  }
\email{zhaoxuzh@usc.edu}

\author{William G. J. Halfond}
\affiliation{%
  \institution{University of Southern California}
  \country{USA}
  }
\email{halfond@usc.edu}

\author{Chunyang Chen}
\affiliation{%
  \institution{Technical University of Munich}
  \country{Germany}
  }
\email{chun-yang.chen@tum.de}

\author{Xiaoxia Sun}
\affiliation{%
  \institution{China Mobile(Suzhou) Software Technology Co., Ltd.}
  \country{China}
  }
\email{18896724798@139.com}

\author{Jiangfan Shi}
\affiliation{%
  \institution{Zhejiang University}
  \country{China}
  }
\email{shijiangfan@dragontesting.cn}

\author{Tingting Yu}
\affiliation{%
  \institution{University of Connecticut}
  \country{USA}
  }
\email{tingting.yu@uconn.edu}

\renewcommand{\shortauthors}{D. Wang, Y. Zhao, S. Feng, Z. Zhang, W. Halfond, C. Chen, X. Sun, J. Shi, and T. Yu}

\begin{abstract}
 In software development, bug report reproduction is a challenging task. This paper introduces \Name{}, a novel feedback-driven approach that leverages GPT-4, a large-scale language model (LLM), to automatically reproduce Android bug reports. Unlike traditional methods, \Name{} bypasses the use of Step to Reproduce (S2R) entities. Instead, it leverages the entire textual bug report and employs innovative prompts to enhance GPT's contextual reasoning. 
 This approach is more flexible and context-aware than the traditional step-by-step entity matching approach, resulting in improved accuracy and effectiveness. In addition to handling crash reports, \Name{} has the capability of handling non-crash functional bug reports.
  Our evaluation of 96 Android bug reports (73 crash and 23 non-crash) demonstrates that 
 \Name{} successfully reproduced 90.63\% of these reports, averaging only 74.98 seconds per bug report. Additionally, \Name{} outperformed three existing tools in both success rate and speed.
\end{abstract} 



\begin{document}
\maketitle
\section{Introduction}
\label{sec:intro}
In software development, debugging and fixing are crucial, especially in the mobile app marketplace. According to~\cite{applause}, 88\% of app users are likely to abandon an app if they encounter recurring issues, underlining the need for swift issue resolution to retain users. One major challenge developers face is effectively reproducing bugs reported by users, which often lack crucial details like the sequence of user interactions~\cite{johnson2022empirical, moran2015auto, bettenburg2008makes, ambriola1997processing}. To address this, the software engineering community is increasingly interested in automating the bug reproduction process.


Several existing approaches have been developed to automate
bug reproduction ~\cite{fazzini2018automatically,zhao2019recdroid, zhao2022recdroid+, zhang2023automatically, feng2024prompting}. These methods  follow two phases in bug reproduction:
1) extracting 
entities from steps to reproduce (S2Rs), and 2) explicitly matching
the extracted entities with the actual app UI to 
find the sequence of events that replays the S2Rs
or reproduces the reported bug.
However, these approaches have limitations. First, S2Rs are often unclear, imprecise, or ambiguous, posing a significant challenge to state-of-the-art NLP techniques~\cite{feng2024prompting}. Second, explicitly matching bug reports with app UI can result in missing reproduction steps due to incomplete bug reports. Existing techniques use resource-intensive dynamic exploration algorithms and human-defined heuristics to address this issue, leading to reduced effectiveness and higher costs.

%


Recent work, 
AdbGPT~\cite{feng2024prompting} utilizes large language models (LLMs), i.e., GPT-3.5, 
to extract S2R entities from bug
reports and then iteratively employs ChatGPT~\cite{IntroducingChatGPT} to make decisions for
selecting UI widgets to replay the
extracted S2Rs. In S2R Entity Extraction, few-shot learning is applied to guide the LLM in recognizing entities related to bug reproduction. The Guided Replay phase uses the LLM to explore the apps, matching S2R entities with GUI events to reproduce the bugs.
This approach leverages the 
remarkable capabilities of
LLMs to comprehend natural language and act as an expert developer
of extracting S2Rs and guiding GUI
exploration.

While AdbGPT represents improvements over previous approaches in effectiveness and efficiency, it still faces several challenges. First, like many existing approaches~\cite{zhao2019automatically, zhao2022recdroid+, zhang2023automatically}, it strictly adheres to the use of S2R entities, following a two-phase structure: the S2R Entity Extraction phase and the subsequent matching with UI widgets. 
As acknowledged in existing work~\cite{moran2015auto, bettenburg2008makes, fazzini2022enhancing}, bug reports often suffer from a substantial cognitive and lexical gap between reporters and developers, leading to ineffective communication of crucial reproduction steps and inconsistent report quality.
The use of S2R entities can exacerbate this situation because (i) the Entity Extraction phase may omit essential details; (ii) the original S2R may be ambiguous or inaccurate; and (iii) extracted entities overlook the actual UI context encountered during bug reproduction.
Second, when implicit input values are involved, AdbGPT uses a “TEST” placeholder for text fields, risking invalid GUI exploration. However, text inputs are crucial for triggering some bugs and significantly influence testing and bug detection.
Third, it overlooks the inherent randomness in large language models' outputs, potentially resulting in inaccuracies in matching GUI widgets and diminishing effectiveness.
Finally, AdbGPT terminates UI exploration when all S2Rs are covered, lacking the capacity to assess whether the bug is being triggered or not.

\ignore{
Although AdbGPT 
represents improvements over previous approaches
in terms of effectiveness
and efficiency, it still
faces several challenges. 
First, similar to many existing approaches~\cite{zhao2019automatically, zhao2022recdroid+, zhang2023automatically}, it still strictly to the use of S2R entities, following a two-phase structure: the S2R Entity Extraction phase and the subsequent matching of these extracted entities with UI widgets. As acknowledged in existing work~\cite{moran2015auto, bettenburg2008makes, fazzini2022enhancing}, bug reports often suffer from a substantial cognitive and lexical gap between reporters and developers, leading to ineffective communication of crucial reproduction steps and inconsistent report quality. The use of S2R entities can exacerbate this situation because (i) the Entity Extraction phase may omit essential details, focusing only on entities like defined user actions and targets; (ii) the original S2R may be ambiguous or inaccurate, with a sole focus on S2R entities further complicating interpretation and accuracy; (iii) the extracted entities, based solely on provided S2Rs, overlook the actual UI context encountered during the real bug reproduction process.
Second, when implicit input values are involved, AdbGPT uses a "TEST" placeholder for text fields, risking invalid GUI exploration due to its less sophisticated input generation approach. However, text inputs are important to trigger some bugs in addition to the
GUI actions, as they significantly influence the testing process and bug detection ~\cite{su2021benchmarking, liu2023fill}.
Third, it overlooks the inherent randomness in Large Language Models' (LLMs) outputs, potentially resulting in inaccuracies in matching GUI widgets and diminishing the effectiveness of finding the correct reproducing sequence. Finally, AdbGPT terminates UI exploration when all S2Rs are covered, focusing primarily on the reply of S2Rs  and lacking the capacity to assess whether the bug is being triggered or not.
}

In this paper, we introduce \Name{}, a novel feedback-driven approach utilizing GPT to automate bug reproduction. Unlike existing methods, \Name{} utilizes the entire bug report, eliminating the need to use S2R entities. This streamlines the process and ensures the original bug report's description remains intact, avoiding potential omissions during the S2R Extraction phase. The feedback-driven design enriches the GPT model with rich UI context, enabling flexible, context-aware actions crucial for accurate bug reproduction.
During reproduction, \Name{} diverges from a rigid step-by-step approach, unlike AdbGPT. Instead, it employs a feedback-driven methodology that considers the bug report, the current app state, and the reproduction history to make informed decisions. \Name{} uses innovative techniques to capture UI context, addressing incomplete and ambiguous bug report information. Additionally, it integrates novel strategies to mitigate randomness in LLMs' responses and automatically adapt to correct behavior.

Overall, \Name{} is a tool that average software developers can easily adopt and benefit from without requiring in-depth knowledge of LLMs technology. First, it is designed to be end-to-end, requiring users to input only the bug report and APK file. Second, it automatically generates feedback based on the bug report, app state, and action status, eliminating the need for manual input. Third, ReBL integrates seamlessly into existing workflows and bug tracking systems, allowing for easy incorporation into bug resolution processes without disruption.


\ignore{

}

\Name{} has been implemented as a powerful software tool built on top of GPT-4~\cite{openai} and UI Automator2~\cite{uiautomator2}. To assess the effectiveness of our approach, we conducted extensive experiments by running \Name{} on a substantial dataset comprising 73 crash bug reports and 23 non-crash bug reports. 
The results show that \Name{} demonstrated an impressive success rate by successfully reproducing 87 bugs (69 crash and 18 non-crash bugs), accounting for 90.63\% of the total bug reports in the dataset, with an average reproduction time of 74.98 seconds. 
To provide further insights into the advantages of our approach over the state-of-the-art, we conducted a comparative analysis of 73 crash bug reports against three existing tools: ReCDroid~\cite{zhao2019recdroid}, ReproBot~\cite{zhang2023automatically}, and AdbGPT~\cite{feng2024prompting}. The success rates for these tools are 45.21\%, 65.75\%, and  73.97\%, respectively, while \Name{} achieved an impressive  94.52\%.
%
Moreover, our approach stands out as the fastest, with an average time of 72.11 seconds, compared to ReCDroid (534.92s), ReproBot (413.72s), and AdbGPT (89.80s).


In summary, our paper makes the following contributions:
\begin{itemize}[topsep=0pt]
\item \Name{}, the first tool capable of reproducing bugs using the whole bug reports without the use of S2R entities and specific bug type domains, streamlining the debugging process.
\item  An empirical study demonstrating the effectiveness of \Name{} in reproducing both crash and non-crash functional bugs for Android bug reports. 
\item We made the implementation and dataset publicly available for future research work~\cite{DEMO}.
\end{itemize}

\begin{figure*}[t]
\centering
\includegraphics[scale=0.21]{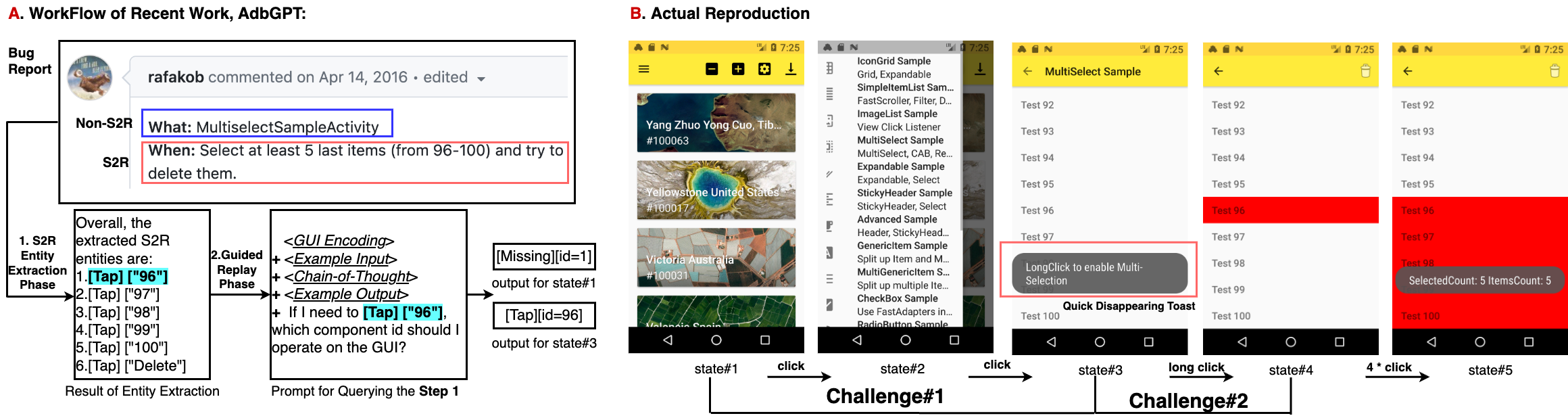}
\caption{Motivation Example}
\label{fig:motivation}
\vspace*{-10pt}
\end{figure*}

\section{Preliminaries and Motivation}

In this section, we introduce the essential preliminaries for automated bug reproduction and provide motivating examples. These examples highlight the limitations of existing approaches and showcase the advantages of our approach.

\subsection{Preliminaries}

A UI  \textit{widget} is a graphical element of an app, such 
as a button, a  text field, and a check box. 
A UI \textit{action} is the action performed by the app. 
It can either be an explicit action on a UI widget (e.g., click) or
an implicit action (e.g., wait, phone call). 
In our setting, 
a \textit{state} represents an app page (i.e., a set of widgets shown on the current screen. If the set of widgets is different, we have another state). 
\emph{UI information} represents the content
of the widgets extracted from the current app state.
\emph{Successful reproduction} is defined as the scenario in which the buggy behavior specified in the bug report is accurately triggered during the bug reproduction process.  



\subsection{Comparison with Existing Techniques}
\label{subsec:moti}
Current state-of-the-art bug report reproduction techniques typically focus on using steps to reproduce (S2Rs) as the initial input for reproduction~\cite{zhao2019automatically, fazzini2018automatically, zhang2023automatically, feng2024prompting}. Some approaches propose automated techniques to extract S2Rs. For example, ReCDRoid+~\cite{zhao2022recdroid+} uses a deep learning algorithm to extract S2Rs
from the full bug reports. 
The extracted S2Rs are typically represented as $<$action, target UI widget, input values$>$.

For the actual bug report reproduction,
existing approaches use various techniques
and algorithms to explicitly match S2R with the 
app UI. This matching process determines the priority of UI widgets in the reproduction approaches' exploration. 
For example, ReCDroid~\cite{zhao2019automatically} 
uses Word2Vec to match S2R entities (e.g., the target UI widget) with the UI widgets in the app and then 
employs a guided DFS to find the most relevant GUI widget iteratively. 
Zhang et al.\cite{zhang2023automatically} calculate
a similarity score to measure the similarity between an action’s UI event and S2R. It then 
uses Q-learning to learn how to match UI events with the S2Rs and bridge missing steps to calculate a UI event sequence that can lead to the observed failure. 
ScopeDroid~\cite{huang2023context} matches the S2R with the state transition graph (STG) generated from the app. 
The matching results are used to plan a path in STG to guide bug report reproduction. 
Despite their contributions, they have inherent limitations, which manifest in at least one of the following aspects: (1) Difficulties in accurately extracting S2R from bug reports, mainly due to the complexity and diversity of sentence structures found in these reports; 
(2) Challenges in inferring missing steps due to the limited knowledge or understanding of the bug reproduction context; 
(3) 
Significant costs associated with the matching and dynamic analysis phases, making the bug reproduction process resource-intensive.



The most recent work, AdbGPT~\cite{feng2024prompting},
addresses the above limitations using large language models (LLMs).
By leveraging their advanced natural language understanding and decision-making capabilities, AdbGPT significantly enhances the accuracy and efficiency of reproducing Android bugs automatically. 
Similar to conventional approaches, AdbGPT initially identifies S2R entities from manually supplied S2R sentences and then use these extracted entities as prompts against the UI widgets to determine the optimal widget for replaying the S2Rs.
Nevertheless, like other techniques, AdbGPT exhibits the following limitations: 



\noindent\textbf{Challenge 1: Overlooking non-S2R information.}
 %
Existing approaches focus on extracting entities from the S2R segment, which risks overlooking other essential information for bug reproduction. 
Figure~\ref{fig:motivation} exemplifies a bug report that failed to be reproduced by existing approaches with S2R entity extraction.
Figure~\ref{fig:motivation}A displays the entity extraction result from AdbGPT~\cite{feng2024prompting}. This result reveals an omission of crucial information, “What:MultiselectSampleActivity”, because it is not in the S2R segment.
The oversight of this non-S2R information results in inadequate information to bridge the missing steps from the home page (state\#1) to this specific page (state\#3), especially since many pages in this app feature a similar UI structure as state\#3, e.g., many pages have items numbered 1-100. Failing to consider this non-S2R information during the reproduction process makes it challenging to determine the exact location where the S2R should be performed. 
Conversely, our approach utilizes the whole bug report and bypasses the Entity Extraction phase, comprehending the bug report as a cohesive whole.

\noindent\textbf{Challenge 2: Incapable of handling incomplete or ambiguous S2Rs.}
%
The S2Rs written by users might be incomplete or ambiguous, and further extraction of S2R entities could potentially exacerbate these issues. Additionally, the extracted entities might lack flexibility, as they are derived from the S2Rs without considering the actual UI context.
%
Continuing with the example from Figure~\ref{fig:motivation},  assume that Challenge 1 has been addressed, allowing AdbGPT to proceed with the reproduction process from state\#3. 
In the Entity Extraction phase, "Select at least 5 last items (from 96-100)" has been broken down by AdbGPT into 5 steps, with the first step being [“tap”] [“96”]. As presented in Figure~\ref{fig:motivation}A, we demonstrate how AdbGPT uses prompt engineering queries in LLMs for suggestions on executing this step. However, while  [“tap”] [“96”] may appear valid according to the bug report and the extraction process, it proves to be invalid in the actual reproduction process. As can be seen in state\#3 in Figure~\ref{fig:motivation}B, a toast, which is a widget that disappears quickly, suggests “LongClick to enable Multi-Selection”, implying that to multi-select items 96-100, one must first \emph{long-click} on item 96. 
Since AdbGPT is only looking to match
the extracted entities $<$[“tap”] [“96”]$>$ to the UI page, 
it will not be able to perform the long
click/tap action. 

To address this problem, our approach eliminates the use of S2R entities, thereby avoiding presuming the action for each step and instead relying on a holistic approach, considering both the complete bug report and the rich UI context to determine the most appropriate action. In this scenario, it recognizes the presence of the quickly disappearing toast message and takes into account its context to perform a long click on the target widget.

\begin{figure}[h]
    \centering
    \begin{subfigure}{0.2\textwidth} 
        \centering
        \includegraphics[width=0.9\linewidth]{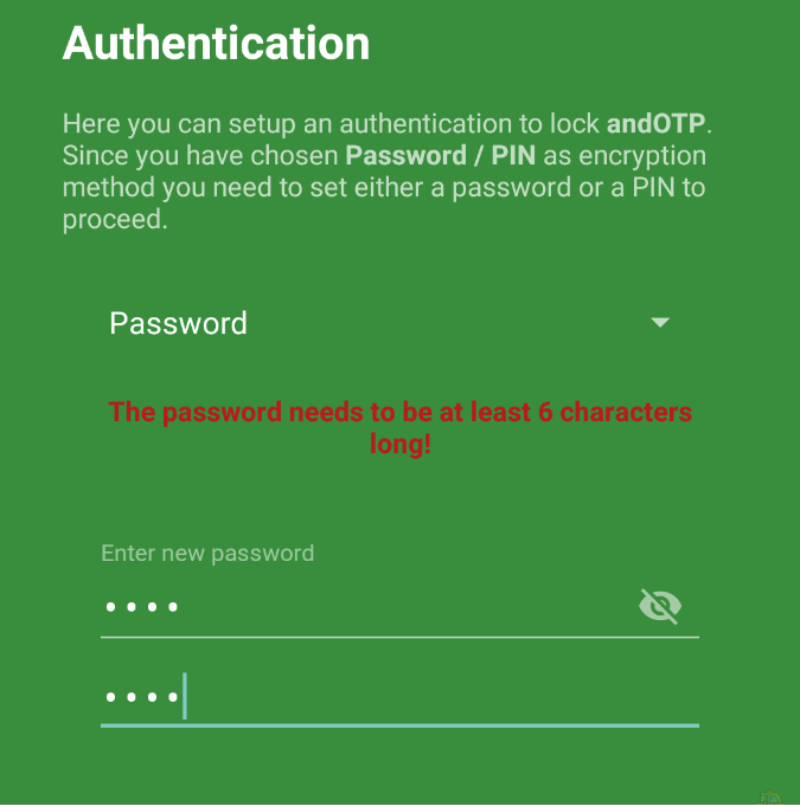} 
        \captionsetup{labelformat=simple, labelsep=period}
        \renewcommand\thesubfigure{\Alph{subfigure}}
        \vspace{-5pt}
        \caption{Fill Blank \#1}
        \label{fig:fill1}
    \end{subfigure}
    \hfill 
    \begin{subfigure}{0.2\textwidth} 
        \centering
        \includegraphics[width=0.9\linewidth]{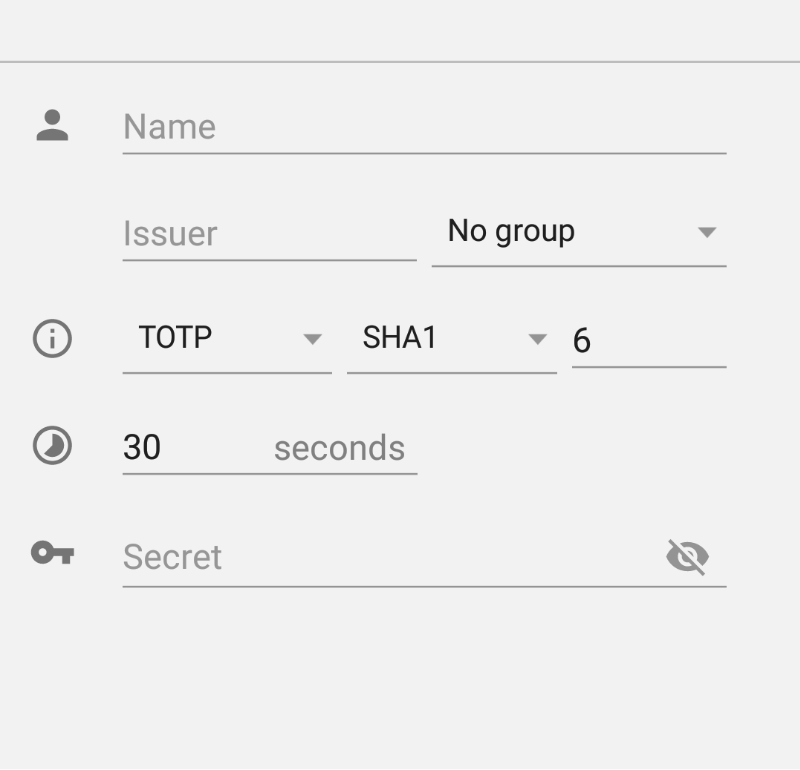} 
        \captionsetup{labelformat=simple, labelsep=period}
        \renewcommand\thesubfigure{\Alph{subfigure}}
        \vspace{-5pt}
        \caption{Fill Blank \#2}
        \label{fig:fill2}
    \end{subfigure}
     \vspace*{-5pt}
    \caption{Example of Inferring Input Values}
    \vspace*{-10pt}
    \label{fig:two_graphs}
\end{figure}

\noindent\textbf{Challenge 3: Less sophisticated text input generation.
}
\label{moti:text}
Existing bug reproduction tools adopt less sophisticated strategies in filling text fields when explicit inputs are lacking in S2Rs, which can lead to invalid inputs, potentially resulting in failed reproduction. These methods include generating random text~\cite{huang2023context}, employing the generic placeholder ~\cite{zhang2023automatically}, relying on predefined dictionaries to fill text boxes, and defaulting to placeholders when unsuitable ~\cite{zhao2019automatically}. AdbGPT~\cite{feng2024prompting} specifically uses the generic placeholder “TEST.”
Particularly challenging are scenarios involving text fields with complex requirements, such as password fields illustrated in Figure~\ref{fig:fill1}. These password fields necessitate a minimum length, and it is common for such fields to demand a combination of letters and numbers, making simplistic placeholders insufficient. Furthermore, password confirmation fields require matching inputs, rendering random generation methods unsuitable. Figure~\ref{fig:fill2} is another example with fields requiring numeric, alphabetic, and unspecified types of input(e.g., the $Secret$ text field) on the same page.  Therefore, a more versatile fill-blank method is necessitated. Addressing these, our approach, \Name{}, intelligently infers input values to fill in blanks leveraging the UI context of the current page. Furthermore, due to the feedback mechanism and flexibility of \Name{}, it is also capable of correcting any invalid inputs, provided there is a warning message, thereby ensuring the accuracy and appropriateness of the inputs.

\noindent\textbf{Challenge 4: Lack of  LLM response management.} 
%
AdbGPT~\cite{feng2024prompting} leverages the advanced capabilities of LLMs but lacks effective mechanisms for managing their outputs. 
When integrating the LLM's output into a program, defining a custom format for automated interpretation is crucial. However, due to the inherent randomness of LLMs, the output might not always adhere to the desired format or could be in the correct format but contain incorrect information, leading to execution failures or program errors. Despite the potential benefits of setting temperature, complete control is not achievable.
Moreover, AdbGPT lacks mechanisms to capture specific output patterns for enhancing bug report reproduction. For instance, the reproduction process may stall when encountering a repeated sequence of actions.
%
%
To address these concerns, our approach uses a feedback mechanism to consistently update the LLMs with feedback on the effectiveness of their responses.
This process guides them toward more accurate and relevant outputs for subsequent actions, thereby enhancing the consistency and reliability of the bug reproduction process.


\noindent\textbf{Challenge 5: Incapable of handling non-crash functional bug reports.}
The wide range of non-crash bug symptoms poses a substantial challenge in bug reproduction. Existing works are primarily focused on crash bug reports ~\cite{fazzini2018automatically,zhao2019automatically, zhao2022recdroid+, zhang2023automatically}, or S2Rs replay without concern for automatically verifying the symptoms, such as AdbGPT~\cite{feng2024prompting}. Crash symptoms are often easy to identify through error messages in Logcat or UI changes, while non-crash bug symptoms are diverse and may need different test oracles for detection.~\cite{xiong2023empirical, wang2022detecting}. Given the advanced text comprehension capabilities of LLMs, we see LLMs' potential to recognize some types of non-crash bug symptoms.
For example, a non-crash symptom described as “See no results” can be effectively determined by LLMs based on the warning “No data” displayed on the screen. Therefore, in this paper, we also investigate whether LLMs can accurately identify content-related non-crash bug symptoms based on the UI context and the symptom described in the bug report, excluding issues related to images such as blurriness, size, and color variations.

\section{\Name{} Approach} 
\label{sec:approach}
\begin{figure}[t]
\centering
\includegraphics[scale=0.29]{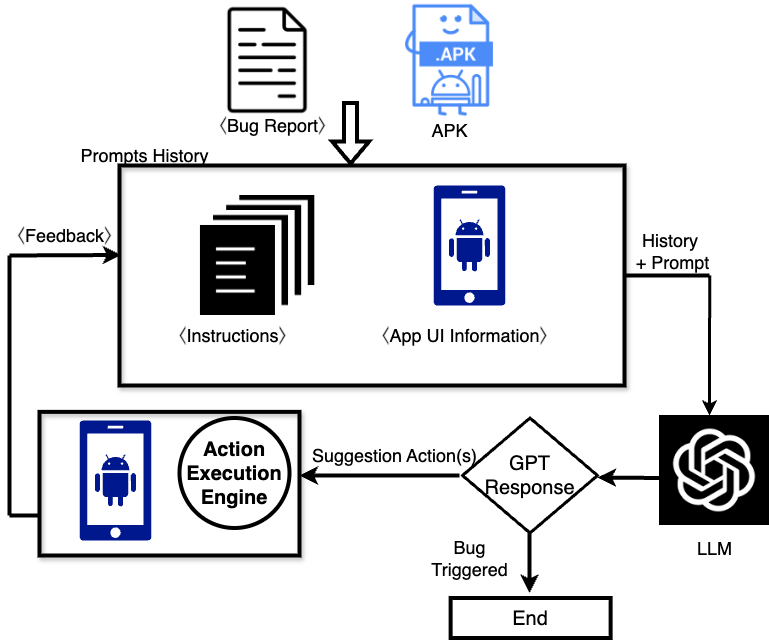}

\caption{ReBL Approach Overview}
\label{fig:Overview}
\end{figure}

%
\Name{} is a feedback-driven approach for automated \textit{whole bug report} reproduction in Android apps utilizing the capabilities of LLMs. The architectural framework of \Name{} is illustrated in Figure~\ref{fig:Overview}. \Name{} is end-to-end, requiring users to input the bug report and APK file. Therefore, developers without knowledge of LLMs can conveniently utilize the tool.
The ultimate objective is to generate an event sequence that precisely reproduces the reported bug. The instructions transform the general-purpose LLM into a bug reproduction tool~\cite{Aligninginstructions, ouyang2022training}, adhering to our design. App UI Information offers the app's UI context. The inherent feedback mechanism of \Name{} is fully automated to generate feedback and provide a richer context of the reproduction process by offering additional observations related to the format of responses, UI context, or actions. 

This is an iterative process. In each iteration, \Name{} leverages the above information to generate prompts and update the prompt history, which are then used to query the LLM for a response. Upon receiving the response, \Name{} interprets it and utilizes the execution engine to perform the suggested actions.  Following this, it updates the feedback and app's state, informing the generation of the next prompt and updating the prompt history. This process continues until the LLM determines the reproduction process should be concluded.

The prompt history preserves all information during the reproduction process, enabling LLMs to maintain a consistent understanding and to reference any detail, such as the bug report and previous UI information, at any time. By utilizing the entire textual bug report, \Name{} bypasses the two traditional phases, S2R entity extraction and S2R entity matching. It directly addresses Challenge\#1. It ensures that every piece of textual information in the bug report is considered. 
Moreover, the description of bug symptoms in the whole bug report combined with the App UI Information aids in determining whether the bug has been triggered, effectively tackling Challenge\#5.

In contrast to AdbGPT, which employs prompts to inquire about precise actions and target S2R entities for each step, our approach takes a significantly different path. \Name{} utilizes the LLM in a distinctive, feedback-driven manner. By providing comprehensive App UI Information and thorough feedback, the LLM is empowered to make well-informed decisions relevant to the current page. This strategy significantly enhances flexibility in bug reproduction, effectively addressing both Challenge\#2, Challenge\#3, and Challenge\#4. Furthermore, in Challenge\#3, if the input for filling a blank generates an invalid response, preventing page progression, the design of feedback can aid in correcting the input. 

\subsection{\textbf{Instructions Description}}
\label{sec:prompt}
The instructions serve as a foundational guide for the LLM in the workflow of automated bug reproduction. Although LLMs possess extensive knowledge, they lack the tailored specialization required for specific tasks such as bug reproduction. The instructions transform the general-purpose LLM into a bug reproduction tool, adhering to the design of our approach. They clearly define the \emph{objective} and \emph{workflow} of the task, followed by a comprehensive \emph{explanation} of the workflow, ensuring that the LLM can perform this specialized task effectively. The structure and components of the instructions  are shown in Table ~\ref{prompt design}.

Equipped with these meticulous prompt instructions, \Name{} acquires the capability to conduct feedback-driven bug report reproduction. This includes supporting advanced actions, executing these actions, gathering UI context, interpreting the LLM response for feedback provision, and establishing criteria for termination. The prompt instructions act as a guiding beacon, steering the entire reproduction process. 

\begin{table}[h]

\begin{threeparttable}
\footnotesize
\caption{Instructions Description}
\label{prompt design}
\centering
\begin{tabular}{p{1.5cm}|p{6.2cm}}
\hline
 \rowcolor{gray!45} \textbf{COMPONENT} & \textbf{DETAILS}\\\hline\hline
⟨Objective⟩ &   I need your assistance in reproducing bug reports for Android apps. Our goal is not just to follow the steps leading to where the bug occurs in the app, but also to verify that the buggy behavior specified in the bug report is indeed triggered.  \\
\hline⟨Workflow⟩ & To initiate the reproduction process, I will provide the app name, bug report, and initial UI information. Your role will be to offer one suggestion at a time, such as clicking a button. After executing your suggestion, I will update you with feedback and the current UI state. This iterative process will continue until either triggering the bug or determining reproduction has failed.
\\
\hline
⟨Explanation⟩ & 
1. Available Actions: click, long\_click,set\_text, scroll,...;
2. Your Response Rormat:...;
3. Termination criteria:
\newline
2. Your response format should be...;\newline
3. The condition to terminate: (a) Successful Reproduction (b) Failed Reproduction.
\\ \hline
\end{tabular}
\begin{tablenotes}
      \item[1]Due to space constraints, this table aims to present the structure and components of the instructions. Full details are available in ~\cite{DEMO}.
\end{tablenotes}
\end{threeparttable}

\vspace{-10pt}
\end{table}


\subsection{Extracting App UI Information}
\label{subsec:ui}
App UI Information showcases the app's current state. It can be used to validate the effectiveness of previous actions and to help make the decision for the next step. In our design, the UI information is composed of two distinct parts: the activity name and the UI widget information.

\subsubsection{Activity Name}
The activity name serves as a unique identifier for the activity within the app. It offers extra context about the functionality or purpose of the current page, aiding in progress checking of bug reproduction.

\subsubsection{UI Widgets Information}
UI widgets showcase the app's current state. 
Existing reproduction approaches~\cite{zhang2023automatically, zhao2019recdroid, zhao2022recdroid+, feng2024prompting} leverage individual UI widgets to perform entity matching between the UI widgets' identifiers (either by their resource-id, content-description, or text) and the extracted S2R entities. We have followed the same method of selecting identifiers to represent single widgets, such as [class: identifier].
If all three identifier fields are empty, then coordinates are used.
However, focusing solely on individual UI widgets can lead to a lack of context, 
making it difficult to manage complex UI pages, such as those with too many widgets or widgets that have identical or semantically similar identifiers on the same page.
It is crucial to group widgets to enhance understanding and facilitate automation tasks~\cite{xie2022psychologically, xiao2024ui}, such as bug reproduction. 

\begin{figure}[h]
  \centering    \includegraphics[scale=0.3]{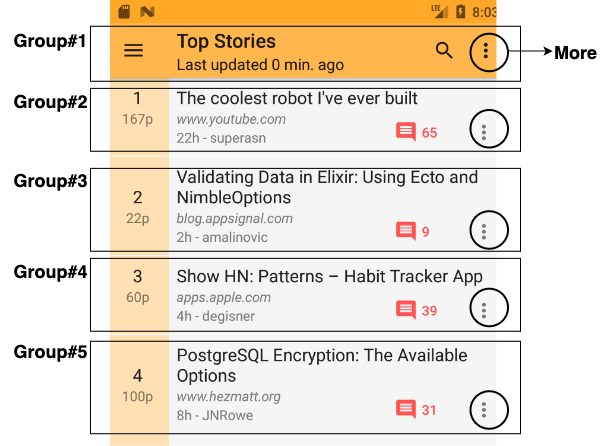} 
    \caption{Illustration of Grouping}
    \label{fig:grouping}
    \vspace*{-10pt}
\end{figure}

Figure~\ref{fig:grouping}  shows an app page featuring many widgets. Viewing these widgets in isolation makes it challenging to discern their specific functions and relationships.  Even for the common widget “More” (known as “more options”), the presence of five such widgets complicates identifying their distinctions when viewed separately.
However, grouping them adds organizational context, which eases the differentiation and prediction of widget relationships and functions. In this example, after grouping, there is a group containing a widget “Top Stories” followed by 
four uniformly structured groups. For instance, each group contains the same number of widgets, including a long text and a URL-like text, suggesting that each group represents a story. The lengthy text likely serves as the title, and the URL-like text acts as the link to that story. Furthermore, one “More” widget in the header group suggests global functionalities for the page, while the other four “More” widgets, distributed among the story groups, indicate local functionalities related to their respective story groups.

\subsubsection{Grouping UI Widgets}
To group UI widgets, 
we analyze the XML of the app page and use the layout structure to systematically group widgets. This approach primarily follows the app developer's intention to organize and contextualize them.
%
In the XML, a \emph{layout} is an element whose class type is ViewGroup or a subclass of ViewGroup, such as LinearLayout or FrameLayout, designed to contain and organize UI components (views) on the screen. 
A \emph{clickable layout} is a layout with its clickable attribute set to true, meaning that 
interacting with any part of the layout triggers a response.  A \emph{child} is an element that is directly contained within another element and is exactly one level below it in the hierarchy. A \emph{leaf} is a child that does not have any children. A \emph{nested layout} is a layout element and serves as a child of another layout element.
Widgets on a page are organized following the format “Group \#[Number]: [List of Widgets]”, where each individual widget within the group maintains the format of a single widget. The rules of grouping work as follows, with the order of steps being critical:

\begin{enumerate}[label=\arabic*.,leftmargin=0.75cm]
    \item For a clickable layout that contains no nested clickable layouts, all widgets within it are considered a group. This rule is adopted because interacting with any part of the layout triggers a response. Widgets within this group collectively convey the group's overall functionality. 
    \item If a non-clickable layout has all its children as leaves, and at least one of them is clickable, all widgets within this layout are considered a group. 
    This rule is based on the understanding that non-clickable widgets can serve as supplementary explanations for clickable widgets in the same group.

    \item If the previous rules do not apply, a clickable widget can be a group by itself. This ensures that widgets not covered by the previous two rules are also taken into consideration.

\end{enumerate}

\begin{figure}[h]
  \centering
    \includegraphics[scale=0.24]{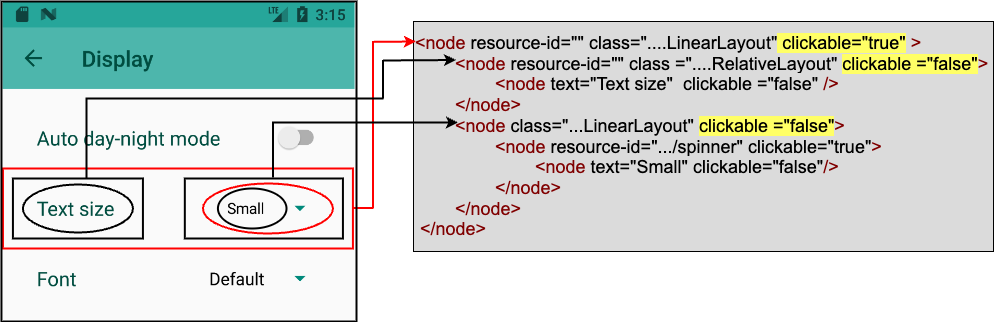} 
    \caption{Grouping Example}
    \label{fig:grouping2} 
    \vspace*{-10pt}
\end{figure}

Figure \ref{fig:grouping2} depicts a UI page for adjusting app display settings. The layout encompassing the “Text size” and “Small” widgets 
constitutes a group because clicking anywhere within this layout prompts the item list for font size. This group is established based on the first rule, as the clickable layout contains no nested clickable layouts. All widgets (i.e., “Text size”, “Small”) within this clickable layout are grouped. 
For the second rule, a common example is a non-clickable label next to an editable text box.

%


%
\subsection{Interpretation and Feedback on LLM Responses}
\label{subsec:response}
\subsubsection{Interpreting the Response}
During each iteration of the bug reproduction process, \Name{} relies on the LLM's response to execute action(s) on the current page. The response can be a single action, such as [$a1$], or a sequence of actions taken in order, such as [$a1$, $a2$, $a3$]. Here, $a$ represents a single action, each paired with its necessary components, such as the target UI widget, input value, direction, and duration. See Table~\ref{Response} for the list of available actions.
%



Compared with existing approaches, \Name{} can handle multiple actions on one page in a single response, which speeds up the reproduction process, saving time from conducting multiple interactions (e.g., sending subsequent prompts and waiting for responses). This proves particularly effective when the bug report requires the selection of multiple items or filling out several text fields on the same page. Moreover, there are scenarios where rapid execution of multiple actions is needed to trigger a bug, such as "Do multiple fast clicks on Play/Stop button"~\cite{transistor-149}. In these instances, existing approaches that execute actions step by step may prove inadequate, while \Name{}, utilizing the multi-actions response from the LLM, is able to handle the rapid execution of a sequence of actions.

\begin{table}[t]
\caption{Actions and LLM
Responses Format}
\vspace*{-10pt}
\centering
\footnotesize
\begin{tabular}{|m{1.5cm}|l|p{3.6cm}|}
\hline
 \rowcolor{gray!45} \multicolumn{3}{|c|}{\textbf{UI Actions}}  \\ \hline \hline
\textbf{Action $a$} &  \textbf{Required Format} & \textbf{Example}\\
\hline
$back$ & [action] & [\textquotesingle back\textquotesingle] \\
\hline
$click$ \newline $long-click$ & [action, target] & [\textquotesingle click\textquotesingle, \textquotesingle theme\textquotesingle]\\
\hline

$scroll$ \newline $swipe$ \newline $rotate$ & [action, direction] & [\textquotesingle scroll\textquotesingle, \textquotesingle up\textquotesingle]\\
\hline
$set\_text$ & [action, target, input] & 
[\textquotesingle set\_text\textquotesingle, \textquotesingle name\textquotesingle, \textquotesingle joh\textquotesingle]\\
\hline \hline
\rowcolor{gray!45} \multicolumn{3}{|c|}{\textbf{System Actions}}  \\ \hline \hline
 $restart$ & [action] & [\textquotesingle back\textquotesingle] \\ \hline
 $sleep$ & [action, duration] & [\textquotesingle sleep\textquotesingle, 0.5]\\
\hline \hline

\rowcolor{gray!45} \multicolumn{3}{|c|}{\textbf{Termination Actions}}  \\ \hline \hline
 $success$& [action] & [\textquotesingle success\textquotesingle] \\ \hline
 $fail$ & [action] & [\textquotesingle fail\textquotesingle] \\ \hline
\hline \hline

 \rowcolor{gray!45} \multicolumn{3}{|c|}{\textbf{LLM Response }}  \\ \hline \hline
Example1 & [a1]&  [[\textquotesingle click \textquotesingle , \textquotesingle A\textquotesingle ]]\\ \hline
Example2  & [a1, a2]&  [[\textquotesingle set\_text\textquotesingle, \textquotesingle email\textquotesingle, \textquotesingle conf@test.com\textquotesingle], [\textquotesingle set\_text\textquotesingle, \textquotesingle password\textquotesingle, \textquotesingle 123456\textquotesingle]]\\ \hline
\end{tabular}
\vspace*{-15pt}
\label{Response}
\end{table}

\subsubsection{Feedback on the Response}
\label{subsec:feedback}
Our approach's design emphasizes the necessity of providing extra feedback on the LLMs' responses at every iteration. This includes (i) execution status, confirming whether the actions were successfully executed; (ii) analysis of whether actions might cause repetition or loops; (iii) observing if actions trigger quick-disappearing widgets that appear and disappear quickly, often unnoticed but might be crucial.

\noindent\textbf{Action execution status.}
The Execution Result handler informs the LLM models of the execution status, indicating whether the previously suggested actions were executed successfully, thereby aiding in the checking of reproduction progress.
Due to the probabilistic nature of language models, the LLM might occasionally produce unexpected responses. For instance, it might suggest performing an action on a UI widget that does not currently exist on the app’s current page, leading to a failure in execution as the target widget cannot be located. Alternatively, the LLM might identify the correct target but output the response in an incorrect format, leading to execution failure as the response cannot be interpreted accurately. To address this, we include the execution result in the  prompts.
This feedback enables the LLM to acknowledge the current status, indicating whether it is appropriate to proceed to the next step or necessary to reformulate the response due to a failed execution. 

\noindent\textbf{Repeated sequence.}
The Repeated Sequence handler detects patterns where a sequence of actions has been executed at least twice, leading to a situation where the reproduction process might get stuck on the same page or enter a loop. 
This handling is particularly crucial because our approach is feedback-driven and does not ask the LLMs to match specific entities with UI widgets. Consequently, there is an inherent potential for the process to become stuck on a page, necessitating the need for this handler to monitor the history of actions, providing crucial oversight to prevent repetitive loops and ensure a smoother reproduction process.
Our algorithm checks if the newly suggested action(s) cause any sequence of actions to repeat from some point until the new actions in each iteration. For example, if the current action history is A → B → C → D → B → C,  and the LLM suggests D as the next action, then a repeated sequence (B → C → D → B → C →D) is detected. When a repeated sequence is detected, we remind the LLM models. This reminder helps the LLM decide if \Name{} should avoid these repetitions in future explorations.


\noindent\textbf{Quick-disappearing widgets.}
A quick-disappearing widget refers to a widget that appears in the UI when it is relevant, often following the execution of an action, but then disappears quickly. Common examples include pop-up notifications, toast, and data-loading dialogs. Existing approaches typically involve a brief waiting period (e.g., 5 seconds) after an action to gather information from the stable UI page, frequently neglecting the existence of quick-disappearing widgets. However, quick-disappearing widgets can be crucial in various aspects of bug reproduction. First, quick-disappearing widgets are crucial for providing information for S2Rs. As exemplified in Challenge 2 in Section~\ref{subsec:moti}, a quick-disappearing toast provides critical information for choosing the correct action. Second, they act as termination indicators, particularly when bug symptoms, such as error messages, are presented as quick-disappearing toast messages. Third, there are situations where the target of reproduction steps is a quick-disappearing widget. To address this challenge, \Name{} is adept at considering the presence and context of quick-disappearing UI widgets, thereby effectively managing scenarios that involve this kind of widget and enhancing the accuracy of the bug reproduction process.

\subsection{Handling Token Limit}
\label{sec:tokenlimit}
The token limit in LLMs restricts the total amount of text that can be included in both the prompt and the model's response, highlighting the necessity for effective management strategies.  
For instance, GPT-4 offers options for specifying token limits --- 8K and 32K, depending on user preference and budget availability~\cite{openai}. In our experiment, we opted for a token limit of 8K.

Our approach employs summarization to address the token limit. With the max token limit ($L$), \Name{} continuously monitors the token count of the prompt history, denoted as $C$. When $C$ surpasses a set threshold ($C > L \times TH$ with $TH$ being the threshold proportion, e.g., 0.7), \Name{} queries the underlying LLM to condense the current prompt history. 
Due to the robust semantic comprehension of LLMs, this method preserves crucial information while significantly reducing the size of prompt history without losing context and eliminating less relevant details for assisting the bug reproduction.
To the best of our knowledge, none of the existing literature has explicitly mentioned specific strategies for addressing the issue of token limits.
Our experiment validates the effectiveness of the summarization strategy.

\subsection{Termination} 
\label{subsec:termination}
The reproduction process terminates, signaling either a successful or failed reproduction.
\vspace*{-5pt}
\subsubsection{Successful Reproduction}
The successful reproduction of a bug report is defined by the manifestation of the reported bug symptom. 
\Name{} leverages the underlying LLM to assess whether a bug has been triggered, utilizing the bug symptom outlined in the report, in conjunction with the current UI information (Section~\ref{subsec:ui}) and the prompt history. The actions executed in the prompt history are used to ensure the mentioned steps in the bug reports have been executed, while the current UI information is employed to verify the presence of bug symptom mentioned in the bug report. For example, \Name{} effectively resolves a \emph{non-crash} bug report reproduction related to an invalid search. 
This was accomplished by utilizing the LLM's capability to  identify the \emph{non-crash} symptom described by the reporter in the bug report as "See no results after search," and subsequently correlating this symptom with the content "No data" displayed on the current UI page. The bug arose due to the expected search yielding no valid results.
This termination strategy also applies to \emph{crash} bugs, where the symptoms are readily apparent.
%
%


%

\subsubsection{Failed Reproduction}
\label{subsub:fail}
Failed reproduction occurs when \Name{} fails to trigger the described bug. In such cases, \Name{} continues to explore the application and makes repeated attempts to reproduce the bug.
This persistent exploration often results in surpassing the token limit of the prompt history as information accumulates over time. Consequently, token limit control mechanisms (Section~\ref{sec:tokenlimit}) are activated.
To optimize resource use and limit endless exploration, failure is determined either after a specified duration (e.g., one hour) or upon reaching a predefined token summarization threshold (e.g., three times).
The one-hour time constraint is aligned with the settings commonly used in state-of-the-art approaches~\cite{zhao2019automatically, zhao2022recdroid+, zhang2023automatically}. Our empirical study sets the token summarization threshold at three because we 
found that typically, one summarization is sufficient to free up space for continued reproduction,  leading to successful outcomes. Therefore, we chose 3 attempts to ensure a sufficient number of opportunities.
For cases that fail to reproduce after three attempts at summarization, the primary reasons for the failures are not due to the token limit, but rather factors such as the underlying tool's capabilities or insufficient information in the original bug report, as detailed in Section~\ref{RQ1}.

\section{Empirical Study}
To evaluate \Name{}, we address three key research questions:

\noindent
{\bf RQ1:} How effective and efficient is \Name{} at reproducing bug reports?

\noindent
{\bf RQ2:} How do individual components contribute to the overall effectiveness and efficiency of \Name{}?

\noindent
{\bf RQ3:}  How does \Name{}'s effectiveness and efficiency in reproducing bug reports compare to that of three baseline approaches?

\subsection{Datasets}
To collect datasets, we adopted the established practice for gathering real-world bug reports for bug reproduction~\cite{fazzini2018automatically, zhao2019recdroid, zhao2022recdroid+,zhang2023automatically, wendland2021andror2, johnson2022empirical}. Specifically, our dataset integrated evaluation datasets from four state-of-the-art tools: AdbGPT~\cite{feng2024prompting}, ReproBot~\cite{zhang2023automatically}, ReCDroid/ReCDroid+~\cite{zhao2019recdroid,zhao2022recdroid+}, and Yakusu~\cite{fazzini2018automatically}, and AndroR2~\cite{wendland2021andror2}, a dataset of manually-reproduced Android bug reports, and an empirical study on Android bug report reproduction ~\cite{johnson2022empirical}. 
We refined our dataset by excluding duplicates, reports related to inaccessible or non-installable APK files, and reports no longer reproducible (e.g., server issues, invalid login). This process resulted in a concise set of 96 unique bug reports, of which 73 are crash reports and 23 are non-crash reports. 
%
%
%
\subsection{Implementation}
We conducted our experiment on a physical x86 machine running Ubuntu 16.04, equipped with an i7-4790 CPU @ 3.60GHz and 32 GB of memory. Notably, this machine did not have a GPU.
For gathering GitHub issues, we utilized the GitHub REST API crawling~\cite{githubapi}. However, for issues present on other platforms like F-Droid, we relied on BeautifulSoup~\cite{beautifulsoup} for data crawling.
Our approach, \Name{}, integrates the underlying language model GPT-4~\cite{openai}.
To facilitate interaction with UI widgets on the device, we implemented UI Automator2~\cite{uiautomator2} as our execution engine.
%
We executed our approach three times to ensure the robustness and consistency of results, and we calculated the average for measuring the execution times.   The implementation of our approach is publicly available, along with the experiment data~\cite{DEMO}.
\subsection{Study Design}
\subsubsection{RQ1: How effective and efficient is \Name{} at reproducing bug reports?}
\label{desing:rq1}
%
Effectiveness is determined by the ratio of successful reproductions to the total number of bug reports examined. 
Efficiency, on the other hand, is measured by the average time taken for successful reproductions. (Refer to Section ~\ref{subsec:termination} for the criteria of successful reproduction.)
The evaluation was conducted within a one-hour timeframe, with the summarization threshold set to three times—a decision explained in Section~\ref{subsub:fail}.
Reproductions that exceeded this time limit or summarization threshold were considered failures. 
To ensure the accuracy of our results and avoid false positives, we conducted a manual inspection to confirm whether the described crash or non-crash bug symptom occurred on the specified target page once  
\Name{} terminated after
executing the given S2Rs in the bug report.

\subsubsection{RQ2: How do the individual components enhance \Name{}'s overall effectiveness and efficiency?} We conducted an ablation study to systematically evaluate the impact of individual components on the functionality and performance of \Name{} by comparing it against a fully functional version. The study included the following three ablations: 1) \textbf{ \Name{}$_{\textbf{S2R}}$} used only S2Rs provided in the bug report, excluding title and other details beyond S2R segment, unlike the full version, which considered the entire textual report;
2) \textbf{ \Name{}$_{\textbf{Indiv}}$} viewed UI widgeted individually, while the full version grouped widgets to process UI information; and 3) 
\textbf{ \Name{}$_{\textbf{NoFB}}$} retained the simple guidelines in the instructions and did not incorporate the feedback mechanism.

\subsubsection{RQ3: How does \Name{}'s effectiveness and efficiency in reproducing bug reports compare to that of three baseline approaches?}
We established three baselines using state-of-the-art automated bug reproduction approaches: AdbGPT~\cite{feng2024prompting}, ReproBot~\cite{zhang2023automatically}, 
and ReCDroid~\cite{zhao2019recdroid}. Throughout our evaluation, we set a time limit of one hour for all techniques to complete the bug reproduction process. This allowed us to assess their performance under consistent conditions and determine their effectiveness and efficiency in reproducing bugs within a reasonable timeframe. Manual inspection was also integral to validate the results, ensuring the reliability of our comparative analysis.

\section{Results and Analysis}
\subsection{RQ1: Effectiveness and Efficiency of \Name{}}
\label{RQ1}

\subsubsection{Effectiveness.}
\label{result:rq1}
\Name{} successfully reproduces 87 out of 96 bugs, including 69 of 73 crash bugs (94.52\%) and 18 out of 23  non-crash functional bugs (78.26\%), achieving an impressive overall success rate of 90.63\%. The details of bug reproductions are shown in \cite{DEMO}. This performance highlights \Name{}'s robustness and versatility in reproducing bug reports. Within the 8k token limit constraint, the summarization mechanism is activated for 11 bug reports to manage token constraints. Among these cases, two~\cite{AIMSICD-816, Fdroidclient-1821} trigger this mechanism, resulting in successful bug reproduction that would otherwise fail. However, the remaining nine reports also activate summarization but fail to reproduce the bug due to reasons other than the token limit.

\noindent \textbf{Reasons for failed reproductions.}
Out of the nine cases where \Name{} fails to reproduce the bugs, four are crash bug reports and five are non-crash bug reports. We identify the following reasons for the failed reproductions in the crash bug reports:

First,  the inability to reproduce bug when   involving third-party services, such as Google Drive.
\Name{} lacks the capability to navigate between different apps, which limits its effectiveness in cases like “ODK-360”~\cite{ODK-360}, where interaction with Google Drive is essential.

Second, the limitation of \Name{}'s underlying testing framework, UI Automator2, appears to be particularly evident in specific apps. It fails to extract custom views from the hierarchy, preventing \Name{} from accessing the necessary UI widgets to reproduce the specified bug. A notable example of this is “Memento-169”~\cite{Memento-169}. Another critical issue is the framework's limited ability to execute certain actions. For example, in “osmeditor-637”~\cite{osmeditor-637}, the framework struggled with  $set\_text$, leading to a failure in the reproduction. These issues may be specific to the compatibility of our underlying testing framework rather than a widespread problem. 

Third, a severe lack of information significantly impedes accurate bug reproduction, as exemplified by the “Anki-6432” case~\cite{Anki-6432}, where the bug report omitted 20 out of the 26 required steps~\cite{zhang2023automatically}. 
\Name{} struggles to identify the bug due to extremely insufficient information in the bug report. Future improvements could include using static analysis~\cite{yang2018static, guo2020improving} to gain comprehensive domain knowledge about the app. This could potentially improve LLMs' understanding and enable more precise predictions, even with limited information in bug reports.

For the five non-crash bug reports, \Name{} faces a unique challenge due to their subtler symptoms compared to crash bugs. This subtlety might lead to false conclusions that a bug has been triggered. For example, in the bug report “LrkFM-34”~\cite{lrkFM-34}, the S2Rs are outlined as follows: “1. Move any file, 2. Try to paste the file, and 3. Observe that nothing happens.”
However, \Name{} fails to reproduce the actual bug because it performs the “move” (essentially cutting a file from one location) and “paste” actions within the same folder. Although this results in “nothing happens” within the same folder -- a symptom that seems to match the bug report -- the actual issue involves failing to paste the item into a different folder, where no files are added after the paste action. 

\begin{figure}[t]
     \vspace{10pt}
      \centering
    \includegraphics[width=0.95\linewidth, height=1.05\linewidth]{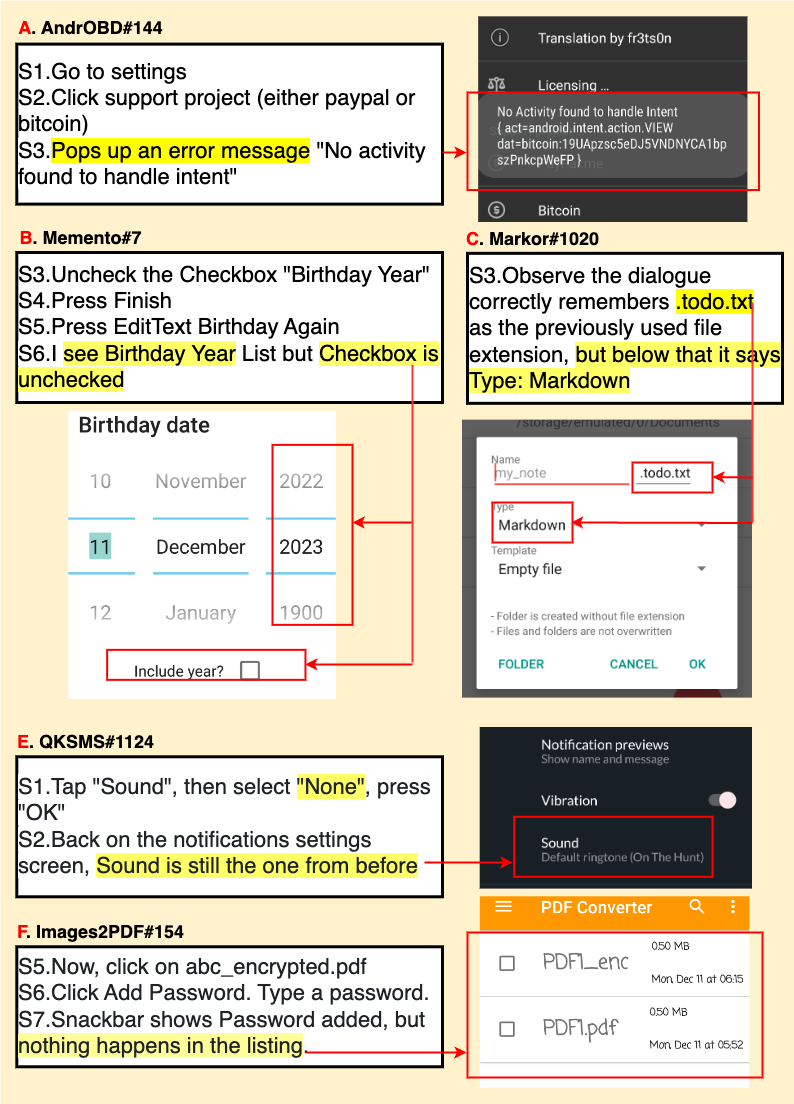}
    \vspace{-10pt}
    \caption{Examples of None-Crash Bug Reports}
      \label{fig:noncrash}
        \vspace{-10pt}
\end{figure}

\noindent \textbf{Successfully reproduced non-crash bug reports.}
 Figure~\ref{fig:noncrash} illustrates some non-crash functional bugs adeptly addressed by \Name{}:

\begin{itemize}[leftmargin=0.25cm]

\item \textbf{Varying warning message.}
Leveraging the semantic capabilities of the underlying LLM, \Name{} excels in \emph{recognizing textual nuances and correlations}. Figure~\ref{fig:noncrash}-A  illustrates an example where the bug report provides an error message. \Name{} identifies that the bug is triggered through the association between the given message and the actual error message. 
Another example, as introduced earlier, \Name{} can associate the “See no results after search” symptom and the “No data” text displayed.

\item \textbf{Missing, redundant, or inconsistent widgets.}
\Name{} can \emph{verify the existence of widgets and their states}. As demonstrated in Figure~\ref{fig:noncrash}-B, \Name{}  determines the presence of the year widget  and verify the state of the checkbox to confirm whether the bug has been triggered. Similarly, Figure \ref{fig:noncrash}-C shows \Name{} handling the inconsistency between displayed type and actual selection.

\item \textbf{Functionality does not take effect.}  
\Name{} 
excels in \emph{comparing changes across pages} by utilizing the historical data in prompt history. Figure \ref{fig:noncrash}-E showcases a bug, the symptom of which is “sound remaining unchanged.” 
\Name{} identifies this bug symptom by accessing and comparing the previous sound settings from the prompt history with the current settings. Likewise, the bug symptom in Figure~\ref{fig:noncrash}-F, “nothing happens in the list,” requires analysis of the UI's state before and after the action. Another common case is the effectiveness of language setting.
%
\end{itemize}

\subsubsection{Efficiency.}     
\Name{} showcases an impressive level of speed in bug reproduction. The time required to reproduce the bugs varied between 19.99 to 243.3 seconds, with a low average time of 74.98 seconds.
 Notably, the bulk of this time is spent interacting with the LLM model, such as making API calls and waiting for responses. In contrast, the processes of action execution, feedback collection, and prompt generation are extremely swift, each taking less than 0.5 seconds.
\begin{tcolorbox}
{\bf RQ1:}  {
\Name{} successfully reproduces 90.63\% of the 96 bug reports with each bug report taking an average of 74.98 seconds. 
It also provides insights into how LLMs can verify non-crash functional bugs, guiding future work in this area.
}
\end{tcolorbox}

\subsection{RQ2: The Roles of Individual Components}

Table~\ref{tab:ablation} shows the results of the three ablations compared with the fully functional \Name{}.

\noindent
\textbf{\Name{}$_{\textbf{S2R}}$} achieves an overall success rate of 81.25\%, including 90.41\% (66/73) for crash reports and  52.17\% (12/23) for non-crash bug reports. This overall success rate is 9.38\% lower than that of \Name{}.
The performance of \Name{}$_{\textbf{S2R}}$ is impacted by the insufficient information when 
considering only S2Rs. 
This shortfall is especially severe for non-crash bug reports because symptoms of non-crash bugs are often found in the title or observed behavior sections, leading to greater oversight. Lacking the symptom of a non-crash bug makes it impossible to verify its occurrence.
\begin{table}[h]
\vspace{-10pt}
\centering
\caption{Ablation study}
\vspace{-10pt}
\label{tab:ablation}
\begin{threeparttable}
\begin{tabular}{|p{1.5cm}|c|c|c|c|}
\hline
 \rowcolor{gray!45}            & \textbf{\small \Name{\_S2R}} & \textbf{\small\Name{\_Indiv}} & \textbf{\small\Name{\_NoFB}} & \textbf{\small\Name{}} \\ \hline\hline
 
Effectiveness    & 81.25\%  &77.08\%  & 73.96\% &  \textbf{90.63\%}   \\ \hline
Efficiency    & 75.50s  & 78.23s  & 87.16 s  & \textbf{74.98s}  \\ \hline
\end{tabular}

\vspace{-10pt}

\end{threeparttable}
\end{table}

\noindent
\textbf{\Name{}$_{\textbf{Indiv}}$} achieves a success rate of 77.08\%, which is 13.55\% lower than \Name{}.
\Name{}$_{\textbf{Indiv}}$ excels when target widget identifiers are clear, similar to existing tools that perform S2R entity match. However, it struggles when the page contains numerous or semantically similar widgets. For example, “open the context menu for an item” targets a group of widgets comprising title and URL, rather than individual ones. Section~\ref{subsec:ui} details limitations regarding individual widgets.

\noindent
\textbf{\Name{}$_{\textbf{NoFB}}$} reports a success rate of 73.96\%, reflecting a reduction of 16.67\% compared to \Name{}.
Its primary challenge is the absence of detailed instructions. Without explanations for actions, \Name{}$_{\textbf{NoFB}}$ may misinterpret actions in the instructions. For instance, “select A” in S2R might be considered a suggestion for “select” rather than “click,” which is the correct action. Furthermore, the lack of feedback exacerbates the issue, as there is no system to indicate failure when attempting to execute “select.”

Efficiency remains consistent across both ablated and fully featured versions, as  the absence of certain details—such as non-S2R information in \Name{}$_{S2R}$ or differences in format—such as  widget format in \Name{}$_{Indiv}$—does not significantly alter prompt size or complexity, thereby not impacting the LLM's processing time. 

%

%

\begin{tcolorbox}
{\bf RQ2:}  {
\Name{} significantly outperformes its three ablated versions, emphasizing the necessity of its full feature set for optimal performance. This highlights the importance of using the whole textual bug report to furnish more comprehensive information, grouping widget to provide structured and organized UI context, providing detailed instructions to ensure smooth interaction with LLMs and implementing a feedback mechanism enables timely measures when unexpected responses occur.
}
\end{tcolorbox}
\subsection{RQ3: Comparison with the State-of-the-Art Approaches} 
\label{rq2}
Table~\ref{tab:combase} presents the overview of the comparison results, comparing our approach with three state-of-the-art baselines across a dataset of 73 crash bug reports. 

\begin{table}[h]
\centering
\vspace*{-10pt}
\caption{Comparison with Baselines}
\label{tab:combase}
\vspace*{-10pt}
\begin{threeparttable}
\begin{tabular}{|l|c|c|c|c|}
\hline
 \rowcolor{gray!45}            & \textbf{\small ReCDroid} & \textbf{\small ReproBot} & \textbf{\small AdbGPT} & \textbf{\small \Name{}} \\ \hline\hline
 
Effectiveness    & 45.21\%  & 65.75\%  & 73.97\% &  \textbf{94.52\%}   \\ \hline
Efficiency    & 534.92s  & 413.72s  &89.80s  & \textbf{72.11s}  \\ \hline
\end{tabular}

\vspace*{-10pt}
\end{threeparttable}
\end{table}

\subsubsection{Effectiveness}
As shown in Table~\ref{tab:combase}, \Name{} successfully reproduces 69 crash bug reports, surpassing the numbers achieved by ReCDroid, ReproBot, and AdbGPT, which reproduce 33, 48, and 54 bug reports, respectively. For bug reports where \Name{} successfully reproduces while the three state-of-the-art tools fail to address, we have conducted a thorough analysis of each case. Our findings reveal four main reasons for these failures, aligning with the challenges we outlined in the motivation section (Section~\ref{subsec:moti}).  It is critical to recognize that it is usually not a single isolated reason leading to the failure, but rather a combination of them.

\noindent \textbf{Reason\#1: Overlooking non-S2R information.} In Challenge 1 (Section~\ref{subsec:moti}), the motivating example shows how the S2R Entity Extraction phase can omit crucial non-S2R information. 
Another example that highlights this issue is \cite{transistor-149}, where one S2R states: “Add URL with the stream”. However, the original post does not contain the specific URL; instead, it is provided as a comment in a different section. Baseline approaches that rely solely on S2R are unable to associate the provided URL.

\begin{figure}[t]
      \centering
        \vspace{10pt}
      \includegraphics[scale=0.3]{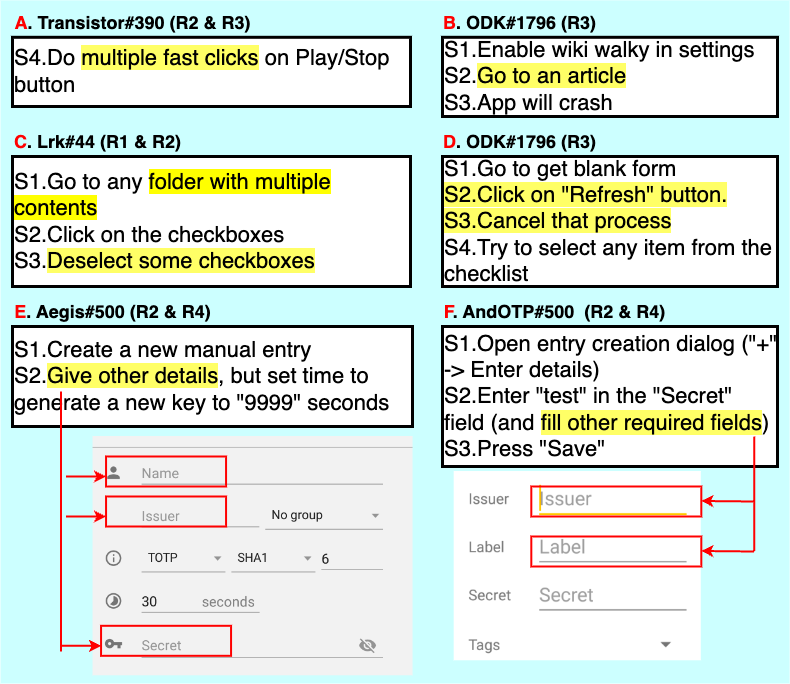}
      \vspace{-10pt}
      \caption{Examples of Other Approaches' Failure Cases}
      \label{fig:examples}
      \vspace{-20pt}
  \end{figure}

\noindent \textbf{Reason\#2: Limitation of S2R entity extraction.}
The S2R Entity Extraction phase sometimes falls short in extracting precise reproduction steps due to its reliance solely on bug report content, neglecting actual context encountered during reproduction. Refer to Figures  \ref{fig:examples}-A and C. They contain ambiguous S2R that make entity extraction challenging. It is challenging to identify the exact number of actions and targets without the actual UI context because “multiple” and “some checkboxes” lack specificity. Similarly, “Given other details” in Figure \ref{fig:examples}-E and “fill other required fields” in Figure \ref{fig:examples}-F face the same limitation.

%

\noindent \textbf{Reason\#3: Challenges in handling incomplete and
ambiguous S2Rs.}
The motivating example in Challenge 2 (Section~\ref{subsec:moti}) underscore this limitation. The experimental results showcase similar instances, primarily attributed to the limited context-awareness of UI information and an inability to handle rapid UI actions.

\noindent{\emph{(i) Limited context-aware:}} 
%
In Figure~\ref{fig:examples}-B, the second step “go to an article” implies the need to leave the settings page and return to the home page to find an article. Existing methods strictly follow sequential steps on the current page, attempting to locate a UI widget within the settings page labeled “article”. However, they may miss the need to navigate elsewhere. If unsuccessful on the current page, these methods typically explore widgets on the same page, rarely considering navigating to a different page unless explicit heuristics guide them. In contrast, \Name{} considers the broader UI context, including the current page and navigational history, while recognizing the current reproduction progress. It prioritizes context-aware actions to achieve the goal of “go to an article,” rather than rigidly focusing on finding a specific S2R entity, “article,” on the current page.
%
\noindent{\emph{(ii) Quick actions:}}
Quick actions are common in using mobile apps and can potentially trigger bugs, as illustrated in Figures \ref{fig:examples}-A and D. In Figure \ref{fig:examples}-A, triggering the bug requires “multiple fast clicks”, presenting a scenario where traditional single-action-per-step approaches fall short. These approaches execute one action per iteration, often followed by a waiting period (e.g., 5 seconds) for UI stabilization. Additionally, processing the next step, such as AdbGPT~\cite{feng2024prompting} querying LLMs, can take 3-10 seconds, and NLP matching approaches also require time. This time gap between actions hinders the triggering of bugs requiring rapid, consecutive interactions, like multiple clicks without interruption.
Figure \ref{fig:examples}-D depicts a concurrency bug where the “refresh” button is part of a rapidly disappearing loading dialog. To trigger this bug, detecting the quick-disappearing widget “Cancel” and successfully clicking on it are necessary, requiring immediate quick action.
Our approach, monitoring UI pages to detect rapidly disappearing widgets and having the flexibility to perform more than one action per step, is well-suited for handling such bug scenarios.

\noindent \textbf{Reason\#4: Fail to generate valid input.}
As noted in Challenge 3 of the motivation section (Section~\ref{subsec:moti}), the inability to generate valid input has been a significant issue in existing works.  Consider Figures \ref{fig:examples}-E and F, where even when the three state-of-the-art tools overcome the above limitation and recognize the need to fill in these blanks, they often provide invalid input. This is primarily because these tools adopt simpler methods for filling text fields when explicit inputs are absent in the S2R. Consequently, they lack the capability to generate context-specific inputs and are unable to correct invalid inputs when warned.

\subsubsection{Efficiency} 
Regarding efficiency, \Name{} demonstrates a significant advantage, with an average reproduction time of 72.11 seconds per bug report. In comparison, ReproBot, ReCDroid, and AdbGPT exhibit considerably longer average times of approximately 413.72 seconds, 534.92 seconds, and 89.80 seconds, respectively. Thus, \Name{} is 7.42 times faster than ReCDroid, 5.74 times faster than ReproBot, and 1.25 times faster than AdbGPT in reproducing bug reports.  


\begin{tcolorbox}
{\bf RQ3:}  {
Our analysis of 73 crash bug reports shows that \Name{} significantly outperforms ReCDroid, ReproBot, and AdbGPT in effectiveness and efficiency. 
Specifically, their success rates are 45.21\%, 65.75\%, and  73.97\% respectively. \Name{} outperforms them with a remarkable success rate of 94.52\%. In terms of efficiency, \Name{} reproduces bug reports in an average time of 72.11 seconds, which is 7.42 times faster than ReCDroid, 5.74 times faster than ReproBot, and 1.25 times faster than AdbGPT in reproducing bug reports. 
}
\end{tcolorbox}

\ignore{
\definecolor{white}{RGB}{255,255,255} 
\definecolor{lightgray}{RGB}{220,220,220} 
\begin{table}
 \scriptsize
  \begin{threeparttable}
  \caption{\commentdbw{will delete the table}}
   \rowcolors{4}{white}{lightgray}
  \begin{tabular}{|p{1.45cm}|c|c|c|c|c|}
    \hline
    \multicolumn{1}{|c|}{} & \multicolumn{1}{c|}{} & \multicolumn{4}{c|}{\textbf{Reproduction Result}}  \\
    \cline{3-6}
      \textbf{Bug Report} & \textbf{Steps} & \textbf{RDroid}& \textbf{RRobot} & \textbf{AdbGPT}   & \textbf{ReBL}\\
    \hline

ActivityDiary\#285 & 0 &\textcolor{red}{$\times$}  & \textcolor{red}{$\times$}  & $\checkmark$&     $\checkmark$ \\\hline
Acv\#12 & 0 & $\checkmark$& $\checkmark$ & $\checkmark$&    $\checkmark$ \\\hline
Aegis\#500 & 0 & \textcolor{red}{$\times$}  &\textcolor{red}{$\times$}  &\textcolor{red}{$\times$}  &     $\checkmark$ \\\hline
Aimsicd\#816 & 0 &\textcolor{red}{$\times$}  & $\checkmark$& $\checkmark$ &      $\checkmark$ \\\hline
AndOPT\#135 & 0 & \textcolor{red}{$\times$}  & $\checkmark$& \textcolor{red}{$\times$} &      $\checkmark$ \\\hline
AndOPT\#500 & 0 &\textcolor{red}{$\times$}   &\textcolor{red}{$\times$}   &\textcolor{red}{$\times$} &     $\checkmark$ \\\hline
AndOPT\#569 & 0 & \textcolor{red}{$\times$} & $\checkmark$&3 &      $\checkmark$ \\\hline
AnglersLog\#9 & 0 & \textcolor{red}{$\times$} &$\checkmark$ & ? &      $\checkmark$ \\\hline
Anki\#4586 & 0 &$\checkmark$ & $\checkmark$& $\checkmark$&     $\checkmark$ \\\hline
Anki\#5638 & 0 & \textcolor{red}{$\times$} &\textcolor{red}{$\times$} &$\checkmark$ &$\checkmark$ \\\hline
Anki\#6432 & 0 & \textcolor{red}{$\times$} & \textcolor{red}{$\times$} &   \textcolor{red}{$\times$} &   \textcolor{red}{$\times$} \\\hline
AntennaPod\#3245$\star$ &0&$\checkmark$  &$\checkmark$ & $\checkmark$ &     $\checkmark$ \\\hline
Anymemo\#18 & 0 & $\checkmark$& $\checkmark$& $\checkmark$&      $\checkmark$ \\\hline
Anymemo\#422$\star$ & 0 & $\checkmark$& $\checkmark$& $\checkmark$ &     $\checkmark$ \\\hline
Anymemo\#440 & 0 &$\checkmark$ & $\checkmark$& $\checkmark$&      $\checkmark$ \\\hline
APhotoMgr\#116 & 0 &$\checkmark$ & $\checkmark$& \textcolor{red}{$\times$}?&     $\checkmark$ \\\hline
AsciiCam\#17 & 0 & \textcolor{red}{$\times$} & \textcolor{red}{$\times$} &3 &     $\checkmark$ \\\hline
Birthdroid\#13 & 0 &$\checkmark$ &$\checkmark$ &$\checkmark$ &     $\checkmark$ \\\hline
Calendula\#134 & 0 & \textcolor{red}{$\times$} & \textcolor{red}{$\times$}& $\checkmark$&     $\checkmark$ \\\hline
CarReport#43 & 0 &$\checkmark$ &\textcolor{red}{$\times$} & $\checkmark$&     $\checkmark$ \\\hline
Commons\#2123 & 0 &$\checkmark$ &$\checkmark$ &\textcolor{red}{$\times$}  &    $\checkmark$ \\\hline
Dagger\#46 & 0 &$\checkmark$ &$\checkmark$ & $\checkmark$ &   $\checkmark$ \\\hline
FamilyFinance\#1 & 0&\textcolor{red}{$\times$} &$\checkmark$ &$\checkmark$ &     $\checkmark$ \\\hline
FastAdapter\#394&    0 & $\checkmark$&$\checkmark$ &  $\checkmark$ &$\checkmark$ \\\hline
FastAdaptor\#113 & 0 & \textcolor{red}{$\times$} & \textcolor{red}{$\times$}&  \textcolor{red}{$\times$}&  $\checkmark$ \\\hline
Fastfitness\#142 & 0 & \textcolor{red}{$\times$} &\textcolor{red}{$\times$}&    $\checkmark$ & $\checkmark$ \\\hline
Fdroid\#1821$\star$ & 0 & \textcolor{red}{$\times$} & \textcolor{red}{$\times$}& $\checkmark$ &      $\checkmark$ \\\hline
Field\#Book\#145 & 0 & \textcolor{red}{$\times$} &$\checkmark$ &\textcolor{red}{$\times$} &    $\checkmark$ \\\hline
Field\#Book\#146 & 0 & \textcolor{red}{$\times$} & \textcolor{red}{$\times$}&$\checkmark$ &    $\checkmark$ \\\hline
FirefoxLite\#5085 & 0 & \textcolor{red}{$\times$} &$\checkmark$ &$\checkmark$ &     $\checkmark$ \\\hline
FlashCards\#13 & 0 & \textcolor{red}{$\times$} &$\checkmark$ & $\checkmark$ &     $\checkmark$ \\\hline
K9\#3255 & 0 &$\checkmark$ &$\checkmark$ &$\checkmark$ &     $\checkmark$ \\\hline
Kiwix\#990$\star$ & 0 &\textcolor{red}{$\times$}  &$\checkmark$ &$\checkmark$ &    $\checkmark$ \\\hline
LibreNews\#22 & 0 &$\checkmark$ &$\checkmark$ &$\checkmark$&     $\checkmark$ \\\hline
LibreNews\#23 & 0 & \textcolor{red}{$\times$} &$\checkmark$ &$\checkmark$ &     $\checkmark$ \\\hline
LibreNews\#27 & 0 & $\checkmark$&$\checkmark$ &\textcolor{red}{$\times$} &     $\checkmark$ \\\hline
Lrk\#44&     0 &\textcolor{red}{$\times$}  &\textcolor{red}{$\times$} & \textcolor{red}{$\times$}&   $\checkmark$ \\\hline
Markor\#1698$\star$ & 0 & \textcolor{red}{$\times$} &$\checkmark$ & $\checkmark$ &  $\checkmark$\\\hline
Markor\#194$\star$& 0 & $\checkmark$&$\checkmark$ &$\checkmark$ &    $\checkmark$ \\\hline
Materialistic\#1067 & 0 &\textcolor{red}{$\times$}  &\textcolor{red}{$\times$} & $\checkmark$ &     $\checkmark$ \\\hline
Memento\#169 & 0 & \textcolor{red}{$\times$} & \textcolor{red}{$\times$}&   \textcolor{red}{$\times$}  &   \textcolor{red}{$\times$}\\\hline
MicroMath\#39 & 0 &\textcolor{red}{$\times$}  &$\checkmark$ & $\checkmark$ &     $\checkmark$ \\\hline
NewsBlur\#1053 & 0 &$\checkmark$ &$\checkmark$ &$\checkmark$  &    $\checkmark$ \\\hline
NoadPlayer#1 & 0 & \textcolor{red}{$\times$} &$\checkmark$ &$\checkmark$  &    $\checkmark$ \\\hline
Notepad\#23 & 0 & $\checkmark$&$\checkmark$ &$\checkmark$ &   $\checkmark$ \\\hline
Obdreader\#22 & 0 &$\checkmark$ &$\checkmark$ &$\checkmark$ &   $\checkmark$ \\\hline
ODK\#360 & 0 &\textcolor{red}{$\times$}  &\textcolor{red}{$\times$} &\textcolor{red}{$\times$}  &\textcolor{red}{$\times$}\\\hline
ODK\#1402$\star$ & 0 &$\checkmark$ &$\checkmark$ &$\checkmark$  &    $\checkmark$ \\\hline
ODK\#1796 & 0 & \textcolor{red}{$\times$} &\textcolor{red}{$\times$} &? &     $\checkmark$ \\\hline
ODK\#2075 & 0 &$\checkmark$ &$\checkmark$ &$\checkmark$ &    $\checkmark$ \\\hline
ODK\#2086 & 0 &$\checkmark$ &$\checkmark$ &$\checkmark$ &     $\checkmark$ \\\hline
ODK\#2191 & 0 & $\checkmark$&$\checkmark$ &$\checkmark$ &    $\checkmark$ \\\hline
ODK\#2525 & 0 & $\checkmark$&$\checkmark$ &$\checkmark$ &    $\checkmark$ \\\hline
ODK\#3222 & 0 &\textcolor{red}{$\times$}  & \textcolor{red}{$\times$}&$\checkmark$ &     $\checkmark$ \\\hline
Olam\#1 & 0 & $\checkmark$&$\checkmark$ &$\checkmark$ &    $\checkmark$ \\\hline
Olam\#2 & 0 & $\checkmark$& \textcolor{red}{$\times$}&$\checkmark$ &      $\checkmark$ \\\hline
Sudoku\#173 $\star$& 0 &$\checkmark$ &$\checkmark$ &$\checkmark$  &    $\checkmark$ \\\hline
Osmeditor\#637 & 0 &\textcolor{red}{$\times$} & \textcolor{red}{$\times$}& $\checkmark$ &\textcolor{red}{$\times$}  \\\hline
PdfViewer\#33 & 0 & \textcolor{red}{$\times$}& \textcolor{red}{$\times$}& 3&     $\checkmark$ \\\hline
Qksms\#482 & 0 &$\checkmark$&$\checkmark$ &$\checkmark$  &    $\checkmark$ \\\hline
Qksms\#585$\star$ & 0 & \textcolor{red}{$\times$}&$\checkmark$ &$\checkmark$ &     $\checkmark$ \\\hline
Screencam\#25 & 0 &$\checkmark$ &$\checkmark$ &$\checkmark$  &    $\checkmark$ \\\hline
Screencam\#32 & 0 & \textcolor{red}{$\times$}&$\checkmark$ &3 &     $\checkmark$ \\\
Soen\#36 & 0 &\textcolor{red}{$\times$} & \textcolor{red}{$\times$}& \textcolor{red}{$\times$}&    $\checkmark$ \\\hline
Timetracker\#10 & 0 &\textcolor{red}{$\times$} &$\checkmark$ & $\checkmark$ &    $\checkmark$ \\\hline
Timetracker\#138 & 0 &\textcolor{red}{$\times$} & \textcolor{red}{$\times$}&\textcolor{red}{$\times$} &    $\checkmark$ \\\hline
Timetracker\#35  & 0 &$\checkmark$ &$\checkmark$ &$\checkmark &  $\checkmark$ \\\hline
Trainer\#7  & 0 & $\checkmark$&$\checkmark$ &\textcolor{red}{$\times$} &    $\checkmark$ \\\hline
Transistor\#149 & 0 &\textcolor{red}{$\times$} & \textcolor{red}{$\times$}&   \textcolor{red}{$\times$}?  &$\checkmark$ \\\hline
Transistor\#63 & 0 & $\checkmark$&$\checkmark$ &$\checkmark$  &    $\checkmark$ \\\hline
Trickytripper\#42 & 0 & \textcolor{red}{$\times$}& \textcolor{red}{$\times$}& $\checkmark$ &     $\checkmark$ \\\hline
Ultrasonic\#187 & 0 & \textcolor{red}{$\times$} &$\checkmark$ &$\checkmark$  &    $\checkmark$ \\\hline
Weather\#61 $\star$ & 0 & $\checkmark$&$\checkmark$ &$\checkmark$ &     $\checkmark$ \\\hline
\textbf{$Success\ Rate$}  & – & \textbf{45.21\%}(33/73) & \textbf{ 65.75\%}(48/73)  & & \\\hline

  \end{tabular}
  \begin{tablenotes}
      \scriptsize
      \item[1] RDroid= ReCDroid\cite{zhao2019automatically}, RRobot= ReproBot\cite{zhang2023automatically}, AdbGPT = AdbGPT\cite{feng2024prompting},  \Name{} = \Name{}. 
      \item[2] $\star$ indicates the bug report include at least one S2R involving rotation
      \item[3] $\checkmark$ = Reproduction Success, \textcolor{red}{\textcolor{red}{$\times$} } = Fail to Reproduce in One Hour. 
      
    \end{tablenotes}
  \label{tab:comparison}
   
  \end{threeparttable}
\end{table}
}
\section{Threats to Validity}
The primary external validity concern in this study revolves around the representativeness of the apps, bug reports, and tools utilized. 
In our evaluation, we aimed to create realistic settings using real bug reports and Android apps. The emulator and execution engine (UI Automator) are widely adopted in both industry and academia, 
consistent with other Android testing works~\cite{wang2023empirical, zhao2022recdroid+, zhang2023automatically,feng2024prompting}.
Furthermore, existing approaches (e.g.,ReCDroid~\cite{zhao2019recdroid} , AdbGPT~\cite{feng2024prompting}]) have demonstrated the effectiveness of such automated reproduction tools over manual reproduction by real-world developers. 
However, we acknowledge that our results may not be fully generalizable to all bug reports in different domains. Additionally, the relatively small number of non-crash bug reports presents an additional constraint, potentially impacting the breadth of our conclusions. 

Regarding internal validity, a notable threat arises from the inherent randomness in the responses generated by LLMs. To address this concern, we ran our experiments three times, thereby reducing the impact of random variations. However, it is essential to recognize that complete consistency in results cannot be guaranteed in all instances. 
\Name{} utilizes GPT-4 as the underlying LLM implementation. While other LLMs~\cite{liu2019roberta, raffel2020exploring,chowdhery2023palm, brown2020language} could be employed, variations in their design and training data may result in different performance outcomes, potentially impacting \Name{}'s effectiveness and accuracy. In the future, we plan to systematically examine the actual impact of various LLMs on our approach. 
\section{Related Work}
\noindent \textbf{Automated Bug Reproduction.}
As discussed in Sections~\ref{sec:intro} and ~\ref{subsec:moti}, there are existing approaches that specifically target automatically reproducing Android bug reports,
including Yakusu~\cite{fazzini2018automatically},  ReCDroid/ReCDroid+~\cite{zhao2019recdroid,zhao2022recdroid+}, MACA~\cite{liuAutomatedClassificationActions2020}, DroidScope~\cite{huang2023context}, and ReproBot~\cite{zhang2023automatically}.
The most recent work,  AdbGPT~\cite{feng2024prompting}, uses LLMs to extract S2R entities for guiding bug report reproduction. However, similar to other existing techniques, AdbGPT's employment of the traditional two-phase structure—consisting of the S2R Entity Extraction phase and the Entity Matching phase—suffers from the same limitations as described in Section~\ref{subsec:moti}.
Our results demonstrate that \Name{} outperforms AdbGPT. 

There are other works that approach bug reproduction from different aspects, such as recording and replaying bugs ~\cite{gomez2013reran, feng2022gifdroid, nurmuradov2017caret, bell2013chronicler, bernal2020translating},  analyzing stack traces~\cite{huang2024crashtranslator}, and leveraging the call stack~\cite{white2015generating}. 
Among these, GIFdroid~\cite{feng2022gifdroid} utilizes screen recordings to automate bug reproduction by adopting image processing techniques. CrashTranslator~\cite{huang2024crashtranslator}  reproduces crash reports directly from stack traces by leveraging a pre-trained LLM to predict the steps necessary for reproduction.

\noindent\textbf{Bug Report Study.}
There have been several research efforts dedicated to studying and analyzing Android bug reports.
For instance, Johnson et al.~\cite{johnson2022empirical} conducted an empirical study on 180 Android bug reports 
to examine their reproduction challenges and the quality of reported details.
Chaparro et al.~\cite{chaparroDetectingMissingInformation2017} conducted an empirical study on user-reported behaviors, reproduction steps, and expected behaviors, identifying discourse patterns used by reporters. Chaparro et al. also developed Euler~\cite{chaparroAssessingQualitySteps2019}, an automatic technique to assess the quality of S2R in Android bug reports, using simple grammar patterns. Liu et al. introduced Maca~\cite{liuAutomatedClassificationActions2020}, a machine learning-based classifier that categorizes action words of S2R into standard categories. However, these techniques all focus on improving the accuracy of identifying S2Rs.

Some research aims to facilitate the reporting process.  For example, Fusion, developed by Moran et al.~\cite{moran2015auto}, employs dynamic analysis to obtain UI events of the app to enhance bug reports during testing. Additionally, Fazzini et al.\cite{fazzini2022enhancing} assist reporters in writing more accurate reproduction steps using information from the static and dynamic analysis of the app to predict the next step. Also, Yang~ et al. ~\cite{song2022toward} provide a guided reporting system with instant feedback and graphical suggestions to improve the quality of bug reports. These approaches improve bug report quality. Though not aimed at reproduction, the improvement in bug report quality could potentially enhance LLMs' comprehension in bug reproduction.

\noindent\textbf{LLMs in Analyzing Bug Reports.} There has been some work on using LLMs to analyze
and understand bug reports. 
Lee et al.~\cite{lee2022light} use LLMs to 
analyze bug reports for bug triage. 
Messaoud et al.~\cite{messaoud2022duplicate}
use the BERT model for duplicate bug report detection. 
Kang et al.~\cite{kang2023large} propose an approach
to use LLMs to generate test methods for Java programs from bug reports. This approach focuses on JUnit tests, which differ from Android UI testing that requires different modeling and iterative exploration processes.

\section{Conclusion}
\label{sec:conclusion}
In conclusion, \Name{} is an advanced automated approach for reproducing both crash and non-crash bug reports in Android apps. Leveraging GPT-4 and well-designed prompts, \Name{} interacts effectively with the GPT model for bug reproduction. Our evaluation, conducted on 96 bug reports from 54 Android apps, showcases \Name{}'s proficiency in successfully reproducing 87 bug reports in an average time of 74.98 seconds, surpassing three existing tools in success rate and efficiency. \Name{} stands out as a lightweight and streamlined solution, offering developers a powerful tool to tackle bug reports efficiently and effectively. In the future, we aim to enhance \Name{}'s performance in handling a wider range of non-crash bug symptoms, potentially through static analysis and fine-tuning. Another goal is to expand \Name{}'s capabilities to tackle more complex and ambiguous S2Rs to solve increasingly intricate scenarios.

\section*{Acknowledgments}
This work was supported in part by the U.S. National
Science Foundation (NSF) under grants 
CCF-2402103, CCF-2403617, CCF-2403747, CCF-2342355, and CCF-2211454.

\bibliographystyle{ACM-Reference-Format}
\bibliography{paper}
\end{document}